\documentclass[prb, twocolumn, superscriptaddress]{revtex4} 
\usepackage{graphicx}
\usepackage{latexsym}
\usepackage{amsmath}
\usepackage{amsthm}
\usepackage{amssymb}
\usepackage{epstopdf} 
\usepackage{enumerate}
\usepackage{setspace} 
\usepackage{dcolumn}
\usepackage{bm}
\usepackage{setspace} 
\usepackage{slashed}
\usepackage{color}
\newcommand{\bk}{{\bm k}}
\newcommand{\bp}{{\bm p}}
\newcommand{\bq}{{\bm q}}

\begin{document}
\title{ 
Long-range Coulomb interaction in nodal ring semi-metals
}

\author{Yejin Huh}
\affiliation{Department of Physics, University of Toronto, Toronto, Ontario M5S 1A7, Canada}

\author{Eun-Gook Moon}
\affiliation{Kadanoff Center for Theoretical Physics and Enrico Fermi Institute, University of Chicago, Chicago, IL 60637, USA}
\affiliation{Department of Physics, Korea Advanced Institute of Science and Technology, Daejeon 305-701, Korea}

\author{Yong Baek Kim}
\affiliation{Department of Physics, University of Toronto, Toronto, Ontario M5S 1A7, Canada}
\affiliation{Canadian Institute for Advanced Research, Toronto, Ontario, M5G 1Z8, Canada}

\date{\today}

\begin{abstract}
Recently there have been several proposals of materials predicted to be nodal ring semi-metals, where zero energy excitations are characterized by a nodal ring in the momentum space. This class of materials falls between the Dirac-like semi-metals and the more conventional Fermi-surface systems. As a step towards understanding this unconventional system, we explore the effects of the long-range Coulomb interaction. Due to the vanishing density of states at the Fermi level, Coulomb interaction is only partially screened and remains long-ranged. Through renormalization group and large-$N_f$ computations, we have identified a non-trivial interacting fixed point. The screened Coulomb interaction at the interacting fixed point is an irrelevant perturbation, allowing controlled perturbative evaluations of physical properties of quasiparticles. We discuss unique experimental consequences of such quasiparticles: acoustic wave propagation, anisotropic DC conductivity, and renormalized phonon dispersion as well as energy dependence of quasiparticle lifetime. 
\end{abstract}

\maketitle

\section{Introduction} 
Tremendous efforts have been made to understand the symmetry-protected gapped topological phases after the discovery of topological insulators \cite{TIrev,SPTrev}. Following this progress, various theoretical and experimental studies have begun to explore the gapless analogs of symmetry-protected topological phases such as the Dirac \cite{neupane14,xu15,liu14} and Weyl semi-metal \cite{wan11,xu15-2,zhang15,xiong15,krempa12}, where low energy excitations possess Dirac-like spectra. Recently, three-dimensional materials with symmetry-protected Fermi line nodes have also been theoretically proposed and experimentally synthesized \cite{burkov11,kee12,yang_2, yige15,schaffer15,rhim15,chen_2,kane15,cava15,mullen15,weng14,yu15,weng15,zeng15}. These systems have nodal rings in momentum space protected by various combinations of time reversal invariance, inversion, chiral and other lattice symmetries. These non-trivial systems are predicted to host topologically protected surface states. However, so far no efforts have been made to study the effects of interactions. 

In this study, we investigate the effects of the long-range Coulomb interaction in nodal ring semi-metals. This is known in various other Fermion systems. In the best-studied system, the Fermi liquid metal, $1/r$-long range interaction is marginal, but Fermi liquid survives due to the strong Thomas-Fermi screening which makes the Coulomb interaction effectively short-ranged. This is caused by metals having an extended Fermi surface and a constant density of states at the Fermi level. The results are known in the other limit, where the energy vanishes only at isolated points of the Brillouin zone. Graphene (in two-dimensions), Weyl semi-metals (in three-dimensions) and double Weyl semi-metals receive logarithmic corrections due to the Coulomb interaction that remains marginal \cite{gonzalez96,neto12,isobe12,lai15,jian15}. In the quadratic band touching case, a non-Fermi liquid phase was found \cite{yang10,moon13}. For anisotropic Weyl fermions, the Coulomb interaction becomes anisotropic and irrelevant \cite{yang14}. 

Nodal ring semi-metals lie in between these two well-studied limits. The energy gap closes on a one-dimensional line node, on which the density of states vanishes. Because of this, short-range interaction was found to be irrelevant \cite{senthil09,lee14}. Screening of the Coulomb interaction is expected to be much weaker compared to the Fermi liquid metal, because fewer states are available to participate. Nonetheless, we show below that the Coulomb interaction is relevant at the non-interacting fixed point. Through renormalization group (RG) analysis and large-$N_f$ computations, we identify a non-trivial interacting fixed point where the partially-screened Coulomb interaction becomes irrelevant, making the fermions asymptotically free in the low energy limit. This allows us to treat the partially-screened Coulomb interaction as a perturbation and calculate the lifetime of the quasiparticles. It is found that the quasiparticle scattering rate vanishes as $E^2$ at low energies even though the partially-screened Coulomb interaction is still long-ranged.

\section{Model}
We start with a non-interacting effective Hamiltonian for the nodal ring semi-metal. This can be written as 
\begin{equation}
\mathcal{H}_0 = \frac{  k_x^2 +k_y^2 -k_F^2 }{2m} \sigma^x+ \gamma k_z  \sigma^y \equiv \epsilon_{a}(k) \sigma^a. \quad a=x,y \;,
\end{equation}
where the Pauli matrices $\sigma_x$ and $\sigma_y$ describe the orbital or pseudo-spin degrees of freedom. This Hamiltonian is similar to that of Ref~\onlinecite{kane15}.
This system has a nodal Fermi ring in the $k_x-k_y$ plane of radius $k_F$, and a linear dispersion in the $k_z$-direction. Its energy spectrum is 
\begin{equation}
E_{\pm}(k)=\pm \sqrt{ \left( \frac{ k_x^2 +k_y^2 -k_F^2}{2m} \right) ^2 + (\gamma k_z)^2}\;,
\label{eq:dispersion}
\end{equation}
for the empty $(+)$ and filled $(-)$ bands. 
In order to describe the effects of Coulomb interaction, we use the Euclidean path integral formalism for the action in $3+1$ dimensions. 
\begin{equation}
\mathcal{S} = \int d\tau d^3 x~\psi^\dagger \left[ \partial_\tau - i e \phi  + \mathcal{H}_0 \right] \psi + \frac{1}{2} \int d\tau d^3 x~ (\partial_i \phi)^2
\label{eq:action1}
\end{equation}
The bosonic field $\phi$ represents the instantaneous Coulomb interaction introduced by the Hubbard-Stratonovich transformation. 

To study how important the interaction is at low energies, we start with finding the engineering dimension of the coupling constant. The non-trivial Fermi surface (ring) in the system affects the scaling dimensions of both fermionic and bosonic fields.

Here we use an RG scheme where a momentum cutoff is applied in the directions around the Fermi ring. We scale the fermion momentum towards the Fermi ring \cite{fitzpatrick13,senthil09}; $k_F$ is fixed and scaling is done only in the Dirac dimensions in which there are linear dispersions. Using definitions $k_r = \sqrt{k_x^2+k_y^2}$ and  $\tilde{k_r} \equiv k_r - k_F$, $\tilde{k_r}$ and $k_z$ are scaled. However there is no scaling in the angular ($\phi \equiv \cos^{-1}(k_x/k_r))$ direction since this represents the gapless degree of freedom. Because of this anisotropy, it is easier to calculate the scaling dimensions from an action written in momentum space rather than in the form given in Eq.~\ref{eq:action1}. Here we generalize the expression to general $d$-spacial dimensions and write the Coulomb interaction as a 4-fermion term. 
\begin{align}
\mathcal{S} & \sim \int_{\omega,\bk}\psi^\dagger (-i\omega + \mathcal{H}_0) \psi \nonumber \\
&~~~+ e^2 \int_{\omega_1,\omega_2,\omega_3,\bk,\bk^\prime, \bq} \frac{1}{q^2} \psi^\dagger(k+q) \psi (k) \psi^\dagger (k^\prime-q) \psi (k^\prime) 
\label{eq:action2}
\end{align}
We have used the notation $\int_\omega = \int d\omega$, $\int_\bk = k_F \int d^{d-1}k \int d\phi$ and $\int_\bq = \int d^{d}q$. 
The constants that have no scaling dimensions such as $k_F$ and $\pi$ have been dropped for clarity. Note that while $k$ and $k^\prime$ are scaled only in the Dirac directions with $d-1$ dimensions, $q$ is scaled in all $d$ dimensions. This is because the important contribution arises from when the momentum carried by the Coulomb interaction is small and when the fermions are close to the Fermi ring. 
The scaling dimensions can be found to be $[\tilde{k_r}] = 1$, $[k_z] = 1$, $[\omega] = 1$, $[q_i]=1$, $[\psi] = -(d+1)/2$, and $[e^2] = 3-d$. Therefore the critical dimension is the physical dimension $d=3$.  
From this we would conclude that the Coulomb interaction to be marginal.  

\section{RG Analysis}
The energy scales of this problem are the Coulomb energy $E_c = e^2 m v_F$, the kinetic energy $E_k = m v_F^2$, and the energy cutoff $E_\Lambda = v_F \Lambda$. We also define a velocity anisotropy parameter $\eta = \gamma / v_F$, where $v_F = k_F/m$ is the fermion radial velocity in the $k_z=0$ plane. The following dimensionless ratios determine the scaling behaviors.  
\begin{align}
\alpha = \frac{E_c}{E_k} = \frac{e^2}{v_F} \quad,\quad 
\beta = \frac{E_c}{E_\Lambda} = \frac{e^2 k_F}{v_F \Lambda} \quad,\quad
\eta = \frac{\gamma}{v_F}
\end{align}
To allow for anisotropic Coulomb interaction, we use as the action for the boson, 
\begin{align}
\mathcal{S}_\phi = \frac{1}{2}\int d\tau d^3 x \left[ a \left( (\partial_x \phi)^2 + (\partial_y \phi)^2 \right) + \frac{1}{a} (\partial_z \phi)^2 \right] \;.
\end{align}
\begin{figure}[]
\includegraphics[width=80mm]{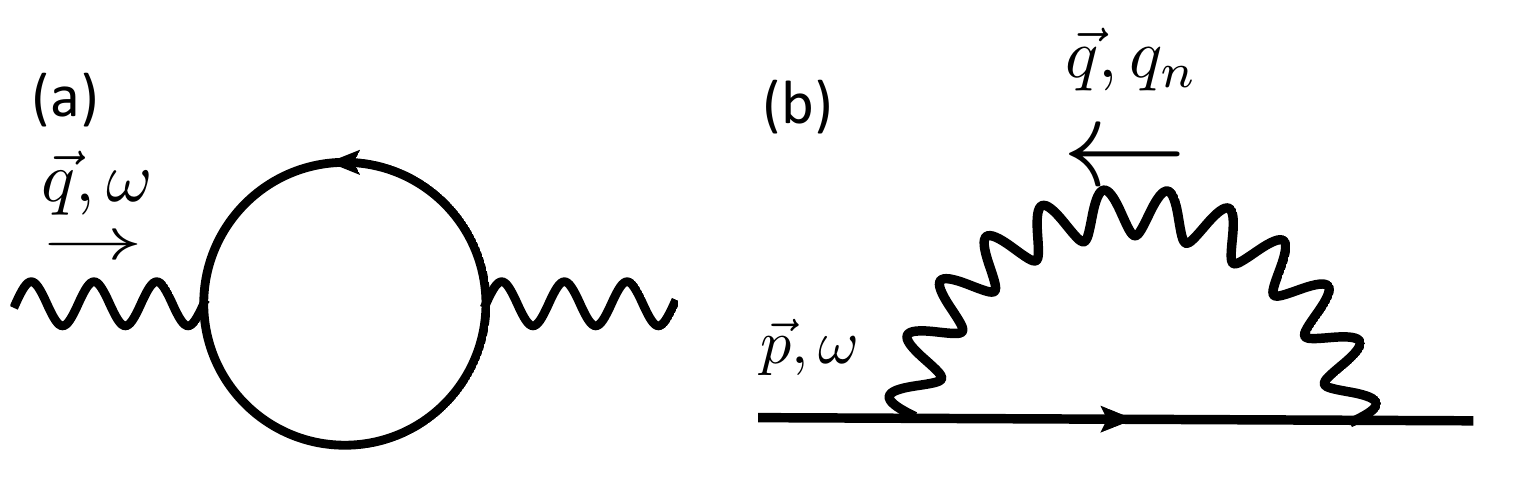}
\caption{Diagrammatic representations of (a) Boson self energy  and (b) Fermion self energy. Straight arrowed lines represent the fermion propagators and the wiggly lines the boson propagators. }
\label{fig:diags1}
\end{figure}
We perform a 1-loop momentum shell RG around the Fermi ring by calculating the boson and fermion self energies to find the RG flow for various parameters. The Feynman diagrams for these self energies are shown in Fig.~\ref{fig:diags1}. 
The boson self energy is, 
\begin{eqnarray}
\Pi(q,i\omega) = -e^2 \int_\bk {\rm Tr}[ G_0(k+q, \Omega+\omega) G_0(k, \Omega) ]\;,
\label{eq:Pi}
\end{eqnarray}
where $G_0(k,i\Omega) = (-i \Omega + \mathcal{H}_0)^{-1}$ is the bare Green's function of the fermions. 

For $\omega =0$, this gives 
\begin{align}
\Pi(\bq, 0) &= e^2 \int_{\bk}^\Lambda \frac{2}{4} \Big(1- \frac{\epsilon_a(k+p/2) \epsilon_a(k-p/2)}{E_{k+q/2} E_{k-q/2}} \Big) \nonumber \\
&~~~~~\times \frac{-2}{ E_{k+q/2} + E_{k-q/2}}\;,
\label{eq:Pi2}
\end{align}
where $E_k=E_+(k)$ is defined by the dispersion relation shown in Eq.~\ref{eq:dispersion}.
We define the momentum shell integral as $\int_\bk^\Lambda = \frac{1}{(2\pi)^3} \int_0^{2\pi} d\phi ( \int_\mu ^\Lambda + \int_{-\Lambda}^{-\mu} ) k_F d \tilde{k_r} \int_{-\infty}^{\infty}dk_z$ with $\mu = \Lambda e^{-d \ell}$. 

The resulting integral can be done after expanding the integrand to second order in $q_r$ and $q_z$. We find 
\begin{align}
\Pi(q_r,q_z) &\approx 
- \frac{e^2}{(2\pi)^2}\left( q_z^2 \frac{ 2 m^2 \gamma}{3 k_F } + q_r^2  \frac{k_F }{ 6\gamma} \right) \frac{d \ell}{\Lambda_r} \nonumber \\
&= -\beta^\prime \left( a q_r^2 \frac{1}{2 a \eta}  + \frac{1}{a}q_z^2 2 a \eta \right) d\ell
 \;,
\label{eq:bosonRG}
\end{align}
where $\beta^\prime = \beta \frac{1}{3(2\pi)^3}$. 
This is infrared (IR) divergent as $\Lambda_r \rightarrow 0$ and the Coulomb interaction is strongly renormalized.

Similarly the fermion self energy is calculated setting external momentum to $\bp = (k_F + p_x, 0, p_z)$. 
\begin{align}
\Sigma_f(\bp) &= -e^2 \int_\bq^\Lambda \frac{\mathcal{H}_0(p+q)}{E_(p_q)} \frac{1}{ a(q_x^2+q_y^2) + 1/a~q_z^2}\nonumber \\
&\equiv -\frac{\alpha}{(2\pi)^2} \left( \sigma_x v_F p_x  F_1(a\eta) + \sigma_y \gamma p_z  F_2(a\eta) \right) d\ell
\end{align}
The momentum shell integral is defined as $\int_\bq^\Lambda = \frac{1}{(2\pi)^3} (\int_\mu ^\Lambda + \int_{-\Lambda}^{-\mu} ) d q_x \int_{-\infty}^{\infty}dq_y dq_z$. The detailed calculation and expressions for $F_1$ and $F_2$ are given in Appendix A. This scaling of the fermion self energy is consistent with the marginal engineering dimension of the bare Coulomb interaction. 
The final RG flow equations for $\alpha$, $\beta^\prime$, and $a \eta$ are
\begin{align}
\frac{d \alpha}{d \ell} &= \alpha \left[ -\frac{1}{2} \beta^\prime \left( \frac{1}{2a\eta} + 2a\eta \right) - \frac{\alpha}{(2\pi)^2} F_1(a\eta) \right]  \nonumber \\
\frac{d\beta^\prime}{d \ell} &= \beta^\prime + \beta^\prime \left[ -\frac{1}{2} \beta^\prime \left( \frac{1}{2a\eta} + 2a\eta \right) -\frac{\alpha}{(2\pi)^2} F_1(a\eta) \right] \nonumber \\
\frac{d(a\eta)}{d \ell} &= a \eta \Bigg[ \frac{1}{2} \beta^\prime \left( \frac{1}{2a\eta} - 2a\eta \right) \nonumber \\
&~~~~~~~~~+ \frac{\alpha}{(2\pi)^2} \left( F_2(a\eta) - F_1 (a\eta) \right) \Bigg]
\end{align}
There are two fixed points: the non-interacting fixed point at $\alpha =0, \beta^\prime = 0$ ($a\eta$ is arbitrary) is unstable and the interacting one at $\alpha = 0, \beta^\prime = 1, a\eta = 1/2$ is stable. From the non-interacting fixed point, $\alpha$ is marginally irrelevant and $\beta$ is relevant. The non-zero value of $\beta^\prime$ at the non-trivial interacting fixed point shows a strong renormalization of the Coulomb interaction while $\alpha=0$ shows that the renormalized Coulomb interaction is irrelevant to the fermions. 

After a step of eliminating high energy degrees of freedom, the boson propagator $D(q)$ can be written as
\begin{align}
&D^{-1}(q) \nonumber \\ 
 &~~= a \left( 1+ \beta^\prime \frac{1}{2a\eta} d\ell \right) (q_x^2+ q_y^2) 
+ \frac{1}{a} \left( 1+ \beta^\prime 2a\eta d\ell \right) q_z^2\;.
\end{align}
Therefore the anomalous dimension is $1$ which arises from the existence of a $k_F$ scale. The renormalized propagator at the new interacting fixed point satisfies 
\begin{align}
D^{-1}(q)&\sim  q_r^{2-1} + |q_z|^{2-1} =  q_r +  |q_z| \;.
\label{eq:ScreenedCoulombRG}
\end{align} 
This will be confirmed by a direct calculation below.

\section{Large $N_f$ calculation}
The screened Coulomb interaction in $d=3$ can also be directly calculated using the random phase approximation. This can be viewed as a large $N_f$ calculation where $N_f$ is the number of fermion flavors. The physical case is $N_f=2$ for the spin states. After introducing a sum over fermion flavors and modifying the coupling constant to $\frac{e}{\sqrt{N_f}}$, the same Eqs.~\ref{eq:Pi} and \ref{eq:Pi2} are calculated without the $\bq$ expansions or the $\bk$ cutoffs. The result is 
\begin{equation}
\Pi(q_r,q_z,\omega=0) = -\frac{e^2}{(2\pi)^3} \left( \frac{k_F q_r}{\gamma} C_1 + 2m |q_z| C_2 \right)\;,
\end{equation}
where $C_1 = 6.86$, $C_2 = 7.28$ are calculated numerically. (Details of the calculation are presented in the Supplemental Material.) 
Therefore for a small $|q|$, the screened Coulomb potential is
\begin{equation}
V_s(q) \sim \frac{1}{\frac{C_1  k_F }{\gamma} q_r+ 2m C_2 |q_z| }\;.
\end{equation}
Notice that the screened Coulomb interaction still has algebraic momentum dependence $1/|q|$ in sharp contrast to that of Fermi liquids. The presence of $k_F$ in nodal ring excitation is not enough to make the Coulomb interaction short-ranged. Furthermore, the directional dependence is qualitatively the same even though the nodal ring spectrum is strongly anisotropic. 
It is important to note that this result is independent of choice in RG scheme since no cutoff has been imposed. The RG calculation is a weak coupling analysis whereas this is a strong coupling analysis with $1/N_f$ as a control parameter. However this result is still consistent with the RG result presented in Eq.~\ref{eq:ScreenedCoulombRG}, which provides validity to both. 

The imaginary part of the bosonic self energy determines the decay. This can be calculated by performing a Wick rotation. This gives the results 
\begin{eqnarray}
{\rm Im} \Pi(q_r=0,q_z,\omega+i0^+) &\sim& \theta (\frac{\omega}{\gamma q_z}-1) \nonumber \\
{\rm Im} \Pi(q_r,q_z=0,\omega+i0^+) &\sim& \frac{\omega^2}{k_F q_r}\;.
\end{eqnarray}
(Full expressions are presented in the Supplmental Material.)
Therefore there is no damping in the direction perpendicular to the ring, while the boson with in-plane momentum shows damping less than that of the Fermi liquid. Landau damping, for comparison, gives ${\rm Im} \Pi(q,\omega) \sim \omega/q$. 

The vertex correction vanishes at the one loop level. This can be easily checked by setting all the external momenta and frequency to $0$. This is as required by the Ward identity because the fermion self energy (Fig.~\ref{fig:diags1}(b)) has no frequency dependence.

\section{Fate of the quasiparticles}
\begin{figure}[]
\includegraphics[width=40mm]{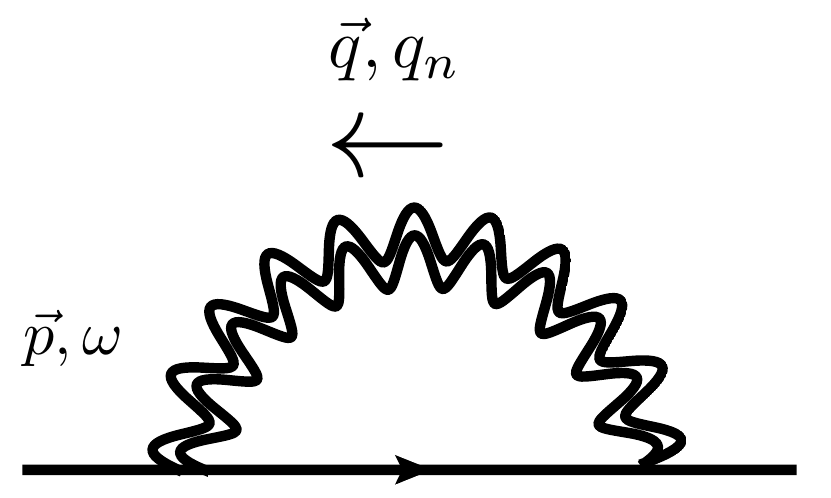}
\caption{ The straight line represents the fermion propagator as in Fig.~\ref{fig:diags1} and the double wiggly line represents the renormalized boson propagator. }
\label{fig:diags2}
\end{figure}
We have seen above that the bosons are strongly renormalized. However since the screened Coulomb interaction is still long-ranged, we must check whether the interaction destroys the fermi liquid or not. The fate of the quasiparticles can be determined from the self energy of the fermions. Using the renormalized boson propagator as shown in Fig.~\ref{fig:diags2}, the self energy is  
\begin{align}
\Sigma_f(p, i \omega_n) &= \int_{\bq, \bq_n} \frac{1}{-i \omega_n - i q_n +\mathcal{H}_0(p+q)} \nonumber \\
&~~~~~~~\times \frac{-e^2}{q^2- \Pi(q, i q_n)} \;.
\end{align}
We find that this is both UV and IR convergent and therefore the screened Coulomb interaction is an irrelevant perturbation to the fermions. Therefore the fermions remain as valid quasiparticles of the system and are effectively decoupled. The result is again consistent with the RG analysis presented earlier.  

The lifetime of these fermions can be found from the imaginary part of the self energy after analytic continuation by the relation $1/\tau = -2{\rm Im}\Sigma_f$. The channel that has the largest contribution is the one that satisfies the Fermi's Golden rule. Focusing on this channel, for a fermion with initial momentum $\bp$ close to the line node and energy $E_p$, we have 
\begin{align}
\frac{1}{\tau} &\sim 2 e^4 \int_{\bk,\bq} \frac{1}{\Pi(q,0)^2} 
\left( 1- \frac{\epsilon_{a}(k) \epsilon_{a}(k+q)}{E_k E_{k+q}}\right) \nonumber \\
&~~~~~~\times \delta( E_{k}+E_{k+q}+E_{p+q}-E_p) \;.
\label{eq:lifetime}
\end{align}
Leading order contributions only come from the region where the intermediate wave vector $\bk$ is very close to the nodal ring. In fact, $\bk$ needs to be closer to the ring than $\bq$ is to the origin. 
We find that $\frac{1}{\tau} \sim \frac{m}{k_F^2}E_p^2 C(\chi_p) $
where $\chi_p$ controls the in-plane component versus the out-of-plane component of $\bp$. $C(\chi_p)$ is a numerical factor that can be numerically calculated for any $\chi_p$. (Details of this calculation can be found in the Appendix.) It is identically zero when the in-plane component disappears. This is because there is no decay channel that satisfies the energy-momentum conservation. This can be seen above by ${\rm Im} \Pi(q_r=0,q_z,\omega) = 0$ when $q_z$ is small. Overall, $1/\tau \sim E_p^2$ and therefore the quasiparticles are long-lived. Interestingly it has the same energy dependence as the Fermi liquid case. While the density of states of this system is vanishing at the Fermi level ($\sim \omega$), this is compensated by the partially screened Coulomb potential. 

%
\section{Experimental signatures} 
Similar to surface acoustic wave propagation experiments in two-dimensional materials \cite{huston62,halperin93,simon96}, a bulk sound wave propagation measurement in a periodic potential with wavelength $\lambda \sim 1/q$ can be used to probe the momentum dependence of the dielectric function. The sound velocity shift and attenuation is found as 
\begin{eqnarray}
\frac{\Delta v_s}{v_s} = \frac{\alpha^2}{2} \frac{1}{1+ f(q)^2}\;, \quad
\kappa = \frac{\alpha^2 q}{2} \frac{ f(q) }{ 1+ f(q)^2}
\end{eqnarray} 
where $f(q) = C \left( \frac{v_s}{v_F} \right)^2 \frac{k_F}{q}$, $C=\frac{\pi}{16}\frac{e^2}{\gamma\epsilon}$ and $\alpha$ is the coupling constant between the piezoelectrics and medium which depends on the geometry.
In contrast, in the Fermi liquid metal, $f(q) = C^{\prime}  \frac{v_s}{v_F}  \left( \frac{k_F}{q} \right)^2$, $C^\prime = \frac{2e^2}{\epsilon v_F}$.

A related physical observable is the DC conductivity. Using the Kubo formula and taking the limits $q \rightarrow 0$, then $\omega \rightarrow 0$, we find the DC conductivity in the clean limit to be finite (due to the underlying particle-hole symmetry) and anisotropic: $\sigma_{xx} = \sigma_{yy} = \frac{e^2}{\hbar}\frac{k_F v_F}{64 \gamma}$ and $\sigma_{zz} = \frac{e^2}{\hbar}\frac{k_F \gamma} {32  v_F}$ with restored units. These results are consistent with a previous compuatation for a similar system \cite{mullen15}. \footnote{In a Weyl semimetal this value approaches 0 linearly as frequency approaches 0 and in fermi liquids this diverges. Another system where a constant value is seen is the two-dimensional graphene.}
The characteristic screening of the Coulomb interaction also affects the phonon dispersion. Longitudinal acoustic phonon dispersion follows $\omega(q)^2 = \Omega_p^2/\epsilon(q)$ where $\Omega_p$ is the plasma frequency of the ions. This shows an unusual $\omega(q) \sim \sqrt{q}$ dependence for small $q$. 

\section{Conclusion}
It is shown that the long-range Coulomb interaction in nodal ring semi-metals leads to a non-trivial fixed point where the screened Coulomb interaction acquires an anomalous dimension. On the other hand, the screened Coulomb interaction becomes irrelevant at the interacting fixed point while remaining long-ranged. Hence the quasi-particles are asymptotically free and physical properties can be computed using a perturbation theory. We show that the quasi-particles have a long life time even though the screening of charged impurity potential would follow an unusual power-law form due to the anomalous dimension. Sound wave propagation and acoustic phonon dispersion show unique momentum dependences. Anisotropic DC conductivity is found and is proportional to the size of the nodal ring. These properties could be tested in future experiments. Interesting future directions include studies of the coupling to critical bosonic modes and impurity/disorder effects. 

\acknowledgments

We thank Jun-Won Rhim, Hae-Young Kee, and Yige Chen for helpful discussions. This work was supported by the NSERC of Canada, Canadian Institute for Advanced Research, and Center for Quantum Materials at the University of Toronto.

\begin{widetext}

\appendix 

\section{RG Calculation}
The fermion self energy is 
\begin{align}
\Sigma_f(k_F + p_x, 0, p_z) &= -e^2 \int_\bq^\Lambda \frac{\mathcal{H}_0(p+q)}{E(p+q)} \frac{1}{ a(q_x^2+q_y^2) + 1/a~q_z^2}\nonumber \\
&\approx -e^2 \int_\bq^\Lambda \left( \frac{ a q_z^2 \gamma^2 }{ ( a^2( q_x^2 + q_y^2 ) + q_z^2 ) ( v_F q_x^2 + \gamma^2 q_z^2 )^ {3/2} } v_F p_x \sigma_x
+ \frac{ a q_x^2 v_F^2 }{ ( a^2( q_x^2 + q_y^2 ) + q_z^2 ) ( v_F q_x^2 + \gamma^2 q_z^2 )^ {3/2} } \gamma p_z \sigma_y \right) 
\end{align}
Using local Cartesian coordinates for $\bq$, we integrate from $-\infty$ to $\infty$ in $q_y$ and $q_z$, and from $\mu = \Lambda e^{-d\ell}$ to $\Lambda$ in $|q_x|$. This gives us

\begin{align}
\Sigma_f(k_F + p_x, 0, p_z) &= -\frac{\alpha}{(2\pi)^2} 
\left( \sigma_x v_F p_x  2 \frac{ a^2 \eta ^2 K\left(1-a^2 \eta ^2\right)-E\left(1-a^2 \eta
   ^2\right)}{ a^2 \eta ^2-1 }
   + \sigma_y \gamma p_z 2 \frac{ \left(E\left(1-a^2 \eta ^2\right)-K\left(1-a^2 \eta
   ^2\right)\right)}{ a^2 \eta ^2-1 } 
 \right)  d\ell \nonumber \\
&\equiv -\frac{\alpha}{(2\pi)^2} \left( \sigma_x v_F p_x  F_1(a\eta) + \sigma_y \gamma p_z  F_2(a\eta) \right) d\ell\;,
\end{align}
where $E(x)$ is the complete elliptic integral of the second kind defined by $E(x) = \int_0^{\pi/2} (1-x\sin^2{\theta})^{1/2} d\theta$ and $K(x)$ is the complete elliptic integral of the first kind defined by $K(x) = \int_0^{\pi/2} (1-x\sin^2{\theta})^{-1/2} d\theta$. The energy ratio $\alpha$ and anisotropy parameter $\eta$ are defined in the main text.

Including the self energies calculated, we can write the effective action as the following. 
\begin{align}
\mathcal{S} &= \mathcal{S} + \int d^4 x \psi^{\dagger}(-\Sigma)\psi + \frac{1}{2} \int d^4x \phi (-\Pi) \phi \nonumber \\
&= \int d^4 x \Bigg[ \psi^\dagger \left( \partial_\tau - i e \phi + \sigma_x v_F \left(1 + \frac{\alpha}{(2\pi)^2} F_1(a\eta) d\ell \right) \partial_r + \sigma_y \gamma \left( 1+ \frac{\alpha}{(2\pi)^2} F_2(a\eta) d\ell \right) \partial_z \right) \psi \nonumber \\
&~~~~~~~~~~~+ 
 \frac{1}{2} a\left( 1 + \beta^\prime \frac{1}{2a\eta} d\ell \right)  \left( (\partial_x \phi)^2 + (\partial_y \phi)^2 \right) + \frac{1}{2 a} \left( 1+ \beta^\prime 2 a \eta d\ell \right)  (\partial_z \phi)^2
  \Bigg]
\end{align}
RG equations for various parameters are 
\begin{align}
\frac{d\ln{v_F}}{d \ell} &= \frac{\alpha}{(2\pi)^2} F_1(a\eta) \nonumber \\
\frac{d\ln{\eta}}{d \ell} &= \frac{\alpha}{(2\pi)^2} \left( F_2(a\eta) - F_1(a\eta) \right) \nonumber \\
\frac{d\ln{e^2}}{d \ell} &= - \frac{1}{2} \beta^\prime \left( \frac{1}{2a\eta} + 2a\eta \right)  \nonumber \\
\frac{d\ln{a}}{d \ell} &= \frac{1}{2} \beta^\prime \left( \frac{1}{2a\eta} - 2a\eta \right)  
\end{align}

Combining these, we can find RG flow equations for $\alpha$, $\beta$, and $a\eta$. These are presented in the main text. 

\section{Large $N_f$ calculation of the screened Coulomb interaction}

By scaling $\tilde{k}_r \equiv k_r - k_F \rightarrow |q_x| r$, $k_F \rightarrow |q_x| \kappa$ and $k_z \rightarrow |q_z| z$, we can write the boson self energy with $\omega = 0$ as the following. We further define $\xi \equiv \frac{q_x^2}{2m} \kappa / (\gamma q_z)$. 

\begin{align}
\Pi(q_x,q_z) &= -\frac{e^2}{(2\pi)^3} \int_{r,\theta,z} k_F |q_x||q_z| 
\left( 1 - \frac{\xi^2(2r +\cos{\theta})(2r - \cos{\theta})+(z+\frac{1}{2})(z-\frac{1}{2})}
{\sqrt{ \xi^2(2r+\cos{\theta})^2 + (z+1/2)^2 } \sqrt{ \xi^2(2r-\cos{\theta})^2 + (z-1/2)^2 }}\right) \nonumber \\
&~~~~ \times \frac{1}{ \sqrt{ \xi^2(2r+\cos{\theta})^2 + (z+1/2)^2 } + \sqrt{ \xi^2(2r-\cos{\theta})^2 + (z-1/2)^2 } } \frac{1}{\gamma |q_z|} \nonumber \\
&= -\frac{e^2}{(2\pi)^3} \frac{k_F |q_x||q_z|}{\gamma |q_z|} f_1(\xi) \nonumber \\
&= -\frac{e^2}{(2\pi)^3} \frac{k_F |q_x||q_z|}{\frac{q_x^2}{2m} \kappa} f_2(\frac{1}{\xi}) 
\end{align}
The second line is better suited to see the behavior of $\xi \gg 1$ and the third line is better suited for $\xi \ll 1$. 
$f_1(\xi)$ can be calculated numerically and fitted by a $C_1 + C_2 \frac{1}{\xi}$ curve which yields $C_1 = 6.86$ and $C_2 = 7.28$. This gives the leading order behavior of the boson self energy at $\omega =0$ provided in the main text.

Effects of finite $\omega$ are seen mainly in the imaginary part of the boson self energy which is $0$ when $\omega = 0$. This can be calculated by performing a standard Wick rotation. 
\begin{eqnarray}
{\rm Im}\Pi(q, \omega + i 0^+)= -\pi e^2 \int_\bk {\rm Tr}\big({\rm P}_{+}(k+q/2) {\rm P}_{-}(k-q/2) \big) \Big(\delta(-\omega + E_{k+q/2} +  E_{k-q/2}) - (\omega \rightarrow -\omega) \Big)
\end{eqnarray} 
${\rm P}_{\alpha}(k)$ are operators that project the states on to the lower and upper bands. 
\begin{equation}
{\rm P}_{\alpha}(k) =\frac{1}{2} \Big(1 + \alpha \frac{\mathcal{H}_{0}(k)}{|E(k)|}\Big)\quad (\alpha=\pm)
\end{equation}
For a positive frequency, the integral becomes 
\begin{eqnarray}
{\rm Im}\Pi(q, \omega + i 0^+)&=& -\pi e^2 \int_\bk {\rm Tr}\big({\rm P}_{+}(k+q/2) {\rm P}_{-}(k-q/2) \big) \delta(-\omega + E_{k+q/2} +  E_{k-q/2}) \nonumber \\
&=& -e^2 \frac{\pi}{(2\pi)^3} \int d k_{\perp} k_{\perp} (2\pi) \int d k_z \frac{2}{4}(1-\frac{\epsilon_a(k+q/2) \epsilon_a(k-q/2)}{E_{k+q/2} E_{k-q/2}}) \delta(-\omega + E_{k+q/2} +  E_{k-q/2})\;.
\end{eqnarray}
The integral is only non-trivial when $\bq \neq 0$. For convenience, we separate it into two cases: one where the external momentum lies in the ring plane and the other where it is perpendicular to the plane. 
\begin{align}
{\rm Im}\Pi(q_z, \omega + i 0^+) 
&= - \frac{\pi e^2}{(2\pi)^2} \frac{1}{8}\frac{m \pi}{\sqrt{ \left( \frac{\omega}{\gamma p_z}\right) ^2-1}} |q_z| \Theta ( \left( \frac{\omega}{\gamma q_z}\right)  - 1 )
\end{align}
\begin{align}
{\rm Im}\Pi(q_x, \omega + i 0^+)=
\begin{cases}
-\frac{e^2}{(2\pi)^3} \frac{k_F |q_x|}{\gamma} \frac{\pi^2}{2}
\left( -E(\frac{\Omega^2}{4}) + K(\frac{\Omega^2}{4})\right)
    & \quad \text{if~} |\Omega| < 2 \\
-\frac{e^2}{(2\pi)^3} \frac{k_F |q_x|}{\gamma} \frac{\pi^2}{4\sqrt{\Omega^2-4}}
\left( -(\Omega^2-4)E(-\frac{4}{\Omega^2-4}) + \Omega^2 K(-\frac{4}{\Omega^2-4})\right)
    & \quad \text{if~} |\Omega| > 2 
\end{cases}\;,
\end{align}
where $\Omega \equiv \frac{2m}{k_F |q_x|} \omega$ is the dimensionless frequency.
$E(x)$ ($K(x)$) is the complete elliptic integral of the second (first) kind defined earlier. 
Asymptotic behavior of this is 
\begin{align}
{\rm Im}\Pi(q_r, \omega + i 0^+)=
\begin{cases}
-\frac{e^2}{(2\pi)^3} \frac{k_F q_r}{\gamma} \frac{\pi^3 \Omega^2}{32} 
	& \quad \text{if~} |\Omega| < 1 \\
-\frac{e^2}{(2\pi)^3} \frac{k_F q_r}{\gamma} \frac{\pi^3}{4 \Omega} 
	& \quad \text{if~} |\Omega| \gg 1 
\end{cases}\;.
\end{align}

\section{Fermion self energy correction}
Here we show the irrelevance of screened Coulomb interaction to the fermions. As a representative example, we only present the renormalization of the fermion dispersion in the $p_z$-direction. For simplicity, we fix the fermion momentum to $\bp = (k_F,0,p_z)$ such that it is only slightly off the line node in the $p_z$-direction and set the frequency to 0. We also fix the internal frequency to 0 (take the Coulomb interaction to be instantaneous) as the effects of a non-zero frequency in the real part of boson propagator is quite small. In the limit where the momentum transfer $|q|$ is small, the bare term of the boson propagator is less important than the self energy correction, and we can set 
\begin{eqnarray}
\Sigma_f(p, 0)  &\rightarrow & \int_{\bq} \frac{ \mathcal{H}_0(p+q) }{ E(p+q) } \frac{ e^2}{N_f\Pi(q, 0)}  \;.
\end{eqnarray}
Here we are only interested in the $\sigma_y$ component of the self energy. Imposing a momentum cutoff in the $q_x$ and $q_y$ direction, we get  
\begin{align}
\Sigma_f(0, p) \sigma_y  & = -\frac{1}{N_f} \gamma p_z \int_{-\Lambda}^\Lambda dq_x \int_{-\Lambda}^\Lambda dq_y \int_{-\infty}^\infty dq_z 
\frac{\left( \frac{1}{2m}(2 k_F q_x + q_x^2 + q_y^2 ) \right) ^2}{ \left(\left( \frac{1}{2m}(2 k_F q_x + q_x^2 + q_y^2 ) \right) ^2 + \gamma^2 q_z^2 \right)^{3/2}}
\frac{1}{C_1 \frac{k_F}{\gamma} |q_y| + C_2 2m |q_z|} \nonumber \\
& \approx  -\frac{1}{N_f}\gamma p_z C_z \frac{\Lambda}{k_F} \;,
\end{align}
where $C_z = \frac{2 G_{4,4}^{4,3}\left(\frac{\text{C1}^2}{4 \text{C2}^2}|
\begin{array}{c}
 \frac{1}{2},1,1,\frac{3}{2} \\
 \frac{1}{2},\frac{1}{2},\frac{1}{2},1 \\
\end{array}
\right)}{\pi ^{3/2} \text{C1}} \sim 1.72$. 
Since the self energy is linear in cutoff, screened Coulomb interaction is irrelevant to the fermions. Alternatively, the full integral without cutoffs can be carried out numerically which gives the same conclusion. 

\section{Lifetime of the fermions}

Starting from the Euclidean fermion self energy, we can calculate the imaginary part of the self energy by analytic continuation. 
\begin{align}
\Sigma_f(p,i\omega ) &= \frac{e^2}{N_f} \int_{\bq,\Omega} \frac{i (\Omega +\omega) + \mathcal{H}(p+q)}{(\Omega +\omega)^2 + E_{p+q}^2} \frac{1}{\Pi(q,\Omega)}\nonumber \\
&\approx -\frac{e^4}{N_f} \int_{\bq, \bk}  
  \frac{2\pi}{\Pi(q,0)^2} \left( 1- \frac{\epsilon_{a}(k) \epsilon_{a}(k+q)}{E_k E_{k+q}}\right)
  \frac{i\omega + \frac{E_{k}+E_{k+q}+E_{p+q}}{E_{p+q}} \mathcal{H}_{p+q}}
  {(E_{k}+E_{k+q}+E_{p+q})^2+\omega^2}
\end{align}
This can be analytically continued by taking $i\omega \rightarrow \omega + i\eta$, and the imaginary part of this would be 
\begin{align}
{\rm Im}\Sigma_f(p,E_p ) &= -\frac{e^4}{N_f} \int_{\bq, \bk} \frac{\pi^2}{\Pi(q,0)^2} 
\left( 1- \frac{\epsilon_{a}(k) \epsilon_{a}(k+q)}{E_k E_{k+q}}\right)
\delta( E_{k}+E_{k+q}+E_{p+q}-E_p)
\left( 1+ \frac{\epsilon_{x}(p+q) }{E_{p+q}} \sigma_x + \frac{\epsilon_{y}(p+q) }{E_{p+q}} \sigma_y \right) \;,
\end{align}
where we have used the approximation
\begin{equation}
\frac{1}{\Pi(q,\Omega)} \approx \frac{\Pi(q,\Omega)}{\Pi(q,0)^2}\;.
\end{equation}
To proceed, we divide the integral into two regions, one where $\bk$ is farther from the Fermi ring than $\bq$ is to the origin ($E_k > E_q$) ($E_q$ is defined in the main text below Eq. 19.) and the opposite case ($E_k < E_q$). For the former case, we can expand the energies up to linear order in $q$. 
\begin{align}
E_{p+q} - E_p &\sim E_q \sin{\chi_q} \sin{\chi_p} + E_q \cos{\chi_q} \cos{\chi_p} \cos{\theta} \nonumber \\
E_k + E_{k+q} &\sim 2 E_k + E_q \sin{\chi_q} \sin{\chi_k} + E_q \cos{\chi_q} \cos{\chi_k} \cos{\phi} \nonumber \\
\delta(E_k + E_{k+q} +E_{p+q} - E_p)
 &\sim \delta\left( 2E_k + E_q \sin{\chi_q} (\sin{\chi_k}+\sin{\chi_p} ) + E_q \cos{\chi_q} ( \cos{\chi_k} \cos{\phi} +\cos{\chi_p} \cos{\theta} ) \right) 
\label{eq:deltaexpand1}
\end{align}
Here we have defined $\phi$, $\theta$, and $\chi_k$ (and similarly $\chi_p$ and $\chi_q$) such that 
\begin{align}
\gamma k_z &= E_k \sin{\chi_k} &   \frac{k_F}{m}\tilde{k}_r &= E_k \cos{\chi_k}  & \vec{k_r}\cdot \vec{q_r} = |k_r||q_r| \cos{\phi} \nonumber \\
\gamma q_z &= E_q \sin{\chi_q} &   \frac{k_F}{m} q_r &= E_q \cos{\chi_q} & \vec{p_r}\cdot \vec{q_r} = |p_r||q_r| \cos{\theta}
\end{align}
where $\vec{k_r}$ is the $\bk$ projected on to the $k_x - k_y$ plane. 
However it is impossible to satisfy the $\delta$-function in Eq.~\ref{eq:deltaexpand1} and $E_k > E_q$ simultaneously. Therefore there is no phase space that conserves momentum and energy in this case, leading to a $0$ contribution to ${\rm Im}\Sigma_f(p,E_p )$. 

For the case where $E_k < E_q$, the second line in Eq.~\ref{eq:deltaexpand1} needs to be modified. Instead of expanding terms in powers of $q$, we expand in powers of $k_r - k_F$ and $k_z$. However $q$ is still assumed to be small compared to $k_F$. 
\begin{align}
E_k + E_{k+q} = E_k  \left( 1 + \frac{\sin{\chi_k} \sin{\chi_q} + \cos{\chi_q} \cos{\chi_k} \cos{\phi}}{ \sqrt{\cos^2{\chi_q} \cos^2{\phi} + \sin^2{\chi_q}}}\right) + E_q \sqrt{ \cos^2{\chi_q} \cos^2{\phi} + \sin^2{\chi_q}  } + O(\frac{E_k^2}{E_q})
\end{align}
In the limit of $E_k / E_q < 1$ the coherence factor becomes, 
\begin{align}
1-\frac{\epsilon_\alpha(k) \epsilon_\alpha(k+q)}{E_k E_{k+q}} = 1-\frac{\sin{\chi_k} \sin{\chi_q} + \cos{\chi_q} \cos{\chi_k} \cos{\phi}}{\sqrt{\cos^2{\chi_q} \cos^2{\phi} + \sin^2{\chi_q}}} + O(\frac{E_k^2}{E_q})
\end{align}
which to leading order, has no energy dependence.
Combining everything, the integral gives
\begin{equation}
{\rm Im} \Sigma_f(p,E_p) = -\frac{\pi^2}{N_f} \frac{4m}{k_F^2} E_p^2 C(\chi_p)\;.
\end{equation}
The largest contributions come from the in-plane scattering where the momentum transfer is opposite to the external momentum. The angle integrals can be done numerically for any given $\chi_p$. For $\chi_p = \pi/2$, which is when the external momentum is only off the Fermi ring in the $z$-direction, this integral is zero, meaning the lifetime is longer than $E_p^2$.

\section{Propagation of Acoustic Waves}

From linear response theory we have, 
\begin{align}
\langle \rho(q,\omega) \rangle = -\chi(q,\omega) \phi_{ext}(q,\omega) = \Pi_0(q,\omega) \phi_{tot}(q,\omega)\;.
\end{align}
The $\Pi_0$ here differs from the RPA calculated in the main text ($\Pi$) by a factor of $(ie)^2$. 

\begin{align}
-\frac{1}{\chi(q,\omega)}&= \frac{1}{\Pi_0(q,\omega)} + V(q) \\
\sigma_{xx} &= -\frac{i \omega}{q^2} \Pi_0(q,\omega) \\
\label{eq:sigma}
-\frac{1}{\chi(q,\omega)}&= \frac{4\pi e^2}{\epsilon q^2 } - \frac{i\omega}{q^2 \sigma_{xx}(q,\omega)}
\end{align}


Using results already obtained, this gives
\begin{align}
-\chi(q=q \hat{z},\omega=v_s q)  
&=  \frac{\epsilon q^2}{4\pi e^2} \frac{1}{1 - i \sigma_m /\sigma_{xx}} 
\end{align}
where $\sigma_m = \frac{\epsilon \omega}{4\pi e^2}$ and 
${\text Re}\sigma_{xx} \approx  \frac{v_s k_F}{64 \gamma} (v_s/v_F)^2$. 

Induced energy per unit area is 
\begin{align}
\delta U &= -\frac{1}{2} \chi |\phi^{ext}|^2 \\
&=  \frac{1}{2} \frac{\epsilon q^2}{4\pi e^2} \frac{1}{1 - i \frac{\sigma_m}{\sigma_{xx}}} |\phi^{ext}|^2 
\end{align}

Measure the energy shift with respect to the shift for $\sigma_{xx} \rightarrow \infty$. 
\begin{align}
\Delta U &= \delta U - \delta U (\sigma_{xx}=\infty) \\
&=  \frac{1}{2} \frac{\epsilon q^2}{4\pi e^2}  \frac{-1}{ 1 - i \frac{ \sigma_{xx} }{ \sigma_{m} } }  |\phi^{ext}|^2 
\end{align}

Acoustic wave with energy density proportional to $q^2$. Therefore the energy $U$ per unit surface area is $U = q^2 C^2 H$. 

\begin{eqnarray}
\frac{\Delta U}{U} = \frac{\Delta q}{q} = -\frac{\Delta v_s}{v_s} + \frac{i \kappa }{q} \\
\frac{\Delta v_s}{v_s} - \frac{i \kappa }{q} =  \frac{\alpha^2/2}{1+i\sigma_{xx}(q,\omega)/\sigma_m}
\end{eqnarray}
This gives the results presented in the main text. 

\end{widetext}


\begin{thebibliography}{}

\bibitem{TIrev}
{M. Z. Hasan and C. L. Kane, Rev. Mod. Phys. {\bf 82}, 3045 (2010); 
X.-L. Qi and S.-C. Zhang, ibid. {\bf 83}, 1057 (2011). }

\bibitem{SPTrev}
{ For a review on recent progress, see T. Senthil, Annual Review of Condensed Matter Physics {\bf 6}, 299 (2015) }

\bibitem{neupane14}
{M. Neupane, S.-Y. Xu, R. Sankar, N. Alidoust, G. Bian, C. Liu, I. Belopolski, T.-R. Chang, H.-T. Jeng, H. Lin, A. Bansil, F. Chou, M. Z. Hasan ,
Nature Commun. 5, 3786 (2014).}

\bibitem{xu15}
{Su-Yang Xu, Chang Liu, Satya K. Kushwaha, Raman Sankar, Jason W. Krizan, Ilya Belopolski, Madhab Neupane, Guang Bian, Nasser Alidoust, Tay-Rong Chang, Horng-Tay Jeng, Cheng-Yi Huang, Wei-Feng Tsai, Hsin Lin, Pavel P. Shibayev, Fang-Cheng Chou, Robert J. Cava, M. Zahid Hasan,
Science 347, 294-298 (2015).}

\bibitem{liu14}
{ Z. K. Liu, B. Zhou, Y. Zhang, Z. J. Wang, H. M. Weng, D. Prabhakaran, S.-K. Mo, Z. X. Shen, Z. Fang, X. Dai, Z. Hussain, Y. L. Chen, 
Science, {\bf 343}, 864 (2014).}

\bibitem{wan11}
{Xiangang Wan, Ari M. Turner, Ashvin Vishwanath, and Sergey Y. Savrasov, 
Phys. Rev. B {\bf 83}, 205101 (2011). }

\bibitem{xu15-2}
{Su-Yang Xu, Ilya Belopolski, Nasser Alidoust, Madhab Neupane, Chenglong Zhang, Raman Sankar, Shin-Ming Huang, Chi-Cheng Lee, Guoqing Chang, BaoKai Wang, Guang Bian, Hao Zheng, Daniel S. Sanchez, Fangcheng Chou, Hsin Lin, Shuang Jia, M. Zahid Hasan, 
 Science 349, 613 (2015). }

\bibitem{zhang15}
{Chenglong Zhang, Su-Yang Xu, Ilya Belopolski, Zhujun Yuan, Ziquan Lin, Bingbing Tong, Nasser Alidoust, Chi-Cheng Lee, Shin-Ming Huang, Hsin Lin, Madhab Neupane, Daniel S. Sanchez, Hao Zheng, Guang Bian, Junfeng Wang, Chi Zhang, Titus Neupert, M. Zahid Hasan, Shuang Jia, 
arXiv:1503.02630 (2015). }

\bibitem{xiong15}
{Jun Xiong, Satya K. Kushwaha, Tian Liang, Jason W. Krizan, Wudi Wang, R. J. Cava, N. P. Ong, 
arXiv:1503.08179 (2015).}

\bibitem{krempa12}
{ W. Witczak-Krempa and Y. B. Kim, 
Phys. Rev. B 85, 045124 (2012). }

\bibitem{burkov11}
{A. A. Burkov, M. D. Hook, and Leon Balents, 
Phys. Rev. B {\bf 84}, 235126 (2011).}

\bibitem{kee12}
{Jean-Michel Carter, V. Vijay Shankar, M. Ahsan Zeb,
and Hae-Young Kee, Phys. Rev. B 85, 115105 (2012).}

\bibitem{yang_2}
{Shengyuan A. Yang, Hui Pan, and Fan Zhang, 
Phys. Rev. Lett. {\bf 113}, 046401 (2015).}

\bibitem{yige15}
{Yige Chen, Yuan-Ming Lu, and Hae-Young Kee, 
Nature Communications {\bf 6}, 6593 (2015).}

\bibitem{schaffer15}
{ R. Schaffer, E. K. H. Lee, Y.-M.Lu, Y. B. Kim, 
Phys. Rev. Lett. {\bf 114}, 116803 (2015).}

\bibitem{rhim15}
{J. W. Rhim and Y. B. Kim,  
Phys. Rev. B {\bf 92}, 045126 (2015) }

\bibitem{chen_2}
{Yuanping Chen, Yuee Xie, Shengyuan A. Yang, Hui Pan, Fan Zhang, Marvin L. Cohen, and Shengbai Zhang, 
Nano Lett. {\bf 15}, 6974 (2015).}

\bibitem{kane15}
{Youngkuk Kim, Benjamin J. Wieder, C. L. Kane, and Andrew M. Rappe, 
Phys. Rev. Lett. 115, 036806 (2015).}

\bibitem{cava15}
{Lilia S. Xie, Leslie M. Schoop, Elizabeth M. Seibel,
Quinn D. Gibson, Weiwei Xie, and Robert J. Cava,
APL Mat. 3, 083602 (2015).}

\bibitem{mullen15}
{Kieran Mullen, Bruno Uchoa, and Daniel T. Glatzhofer, 
Phys. Rev. Lett. {\bf 115} 026403 (2015).}

\bibitem{zeng15}
{Minggang Zeng, Chen Fang, Guoqing Chang, Yu-An Chen, Timothy Hsieh, Arun Bansil, Hsin Lin, and Liang Fu, 
arXiv:1504.03492 (2015).}

\bibitem{weng14}
{ Hongming Weng, Yunye Liang, Qiunan Xu, Yu
Rui, Zhong Fang, Xi Dai, Yoshiyuki Kawazoe,
Phys. Rev. B 92, 045108 (2015).}

\bibitem{yu15}
{Rui Yu, Hongming Weng, Zhong Fang, Xi Dai, Xiao Hu,
Phys. Rev. Lett. 115, 036807 (2015) .}

\bibitem{weng15}
{Hongming Weng, Chen Fang, Zhong Fang, B. Andrei
Bernevig, and Xi Dai, Phys. Rev. X 5, 011029 (2015) .}

\bibitem{neto12}
{Valeri N. Kotov, Bruno Uchoa, Vitor M. Pereira, F. Guinea, and A. H. Castro Neto, 
Rev. Mod. Phys., {\bf 84}, 1067 (2012). }

\bibitem{gonzalez96}
{Gonzalez, J., F. Guinea, and M. A. H. Vozmediano, 
Phys. Rev. Lett. {\bf 77}, 3589. (1996). }

\bibitem{isobe12}
{Hiroki Isobe and Naoto Nagaosa, 
Phys. Rev. B. {\bf 86}, 165127 (2012).}

\bibitem{lai15}
{Hsin-Hua Lai, 
Phys. Rev. B  {\bf 91}, 235131 (2015).  }

\bibitem{jian15}
{Shao-Kai Jian and Hong Yao, 
Phys. Rev. B.  {\bf 92}, 045121 (2015).}

\bibitem{yang10}
{B. J. Yang and Y. B. Kim, 
Phys. Rev B. {\bf 82}, 085111 (2010).}

\bibitem{moon13}
{Eun-Gook Moon, Cenke Xu, Yong Baek Kim, and Leon Balents,
Phys. Rev. Lett. {\bf 111} 206401 (2013). }

\bibitem{yang14}
{Bohm-Jung Yang, Eun-Gook Moon, Hiroki Isobe, and Naoto Nagaosa, 
Nat. Phys. {\bf 10} 774 (2014). }

\bibitem{senthil09}
{T. Senthil and R. Shankar, Phys. Rev. Lett. {\bf 102} 046406 (2009).}

\bibitem{lee14}
{E. K. H. Lee, S. Bhattacharjee, K. Hwang, H.-S.Kim. H. Jin, Y. B. Kim, 
Phys. Rev. B. {\bf 89}, 205132 (2014).}

\bibitem{fitzpatrick13}
{A. Liam Fitzpatrick, Shamit Kachru, Jared Kaplan, and S. Raghu, 
Phys. Rev. B {\bf 88} 125116 (2013). }

\bibitem{huston62}
{A. R. Huston and Donald L. White, 
J. Appl. Phys. {\bf 33}, 40 (1962). }

\bibitem{halperin93}
{B. I. Halperin, Patrick A. Lee, and Nicholas Read,
Phys. Rev. B. {\bf 47} 7312 (1993). }

\bibitem{simon96}
{Steven H. Simon, 
Phys. Rev. B. {\bf 54} 13878 (1996). }



\end{thebibliography}
\end{document}